\documentclass[12pt]{article}

\usepackage{epsfig}

\def\be{\begin{equation}}
\def\ee{\end{equation}}

\def\IP{\hbox{\rm I\kern -1.6pt{\rm P}}}
\def\IC{{\hbox{\rm C\kern-.58em{\raise.53ex\hbox{$\scriptscriptstyle|$}}
    \kern-.55em{\raise.53ex\hbox{$\scriptscriptstyle|$}} }}}
\def\IN{\hbox{I\kern-.2em\hbox{N}}}
\def\IR{\hbox{\rm I\kern-.2em\hbox{\rm R}}}
\def\ZZ{\hbox{{\rm Z}\kern-.3em{\rm Z}}}
\def\IT{\hbox{\rm T\kern-.38em{\raise.415ex\hbox{$\scriptstyle|$}} }}

\begin{document}

\title{On the complexity of curve fitting algorithms}
\author{N. Chernov, C. Lesort, N. Sim\'{a}nyi\\
Department of Mathematics\\
University of Alabama at Birmingham\\
Birmingham, AL 35294, USA}
\date{\today}
\maketitle




\begin{abstract}
We study a popular algorithm for fitting polynomial curves to scattered
data based on the least squares with gradient weights. We show that
sometimes this algorithm admits a substantial reduction of complexity,
and, furthermore, find precise conditions under which this is possible.
It turns out that this is, indeed, possible when one fits circles but
not ellipses or hyperbolas.
\end{abstract}





In many applications one needs to fit a curve described by a polynomial
equation
$$
           P(x,y;\Theta)=0
$$
(here $\Theta$ denotes the vector of unknown parameters) to
experimental data $(x_i,y_i)$, $i=1,\ldots,n$. In this equation $P$ is
a polynomial in $x$ and $y$, and its coefficients are either unknown
parameters or functions of unknown parameters. For example, a number of
recent publications \cite{CBH01,GGS94,LM00} are devoted to the problem
of fitting quadrics $Ax^2+ Bxy+ Cy^2+ Dx+ Ey+ F=0$, in which case
$\Theta=(A,B,C,D,E,F)$ is the parameter vector. The problem of fitting
circles, given by equation $(x-a)^2+ (y-b)^2 -R^2=0$ with three
parameters $a,b,R$, also arises in practice \cite{CO84,Ka98}.

It is standard to assume that the data $(x_i,y_i)$ are noisy
measurements of some true (but unknown) points $(\bar{x}_i,\bar{y}_i)$
on the curve, see \cite{BC86,CL03a,Ka96,Ka98} for details. The noise
vectors $e_i=(x_i-\bar{x}_i,y_i-\bar{y}_i)$ are then assumed to be
independent gaussian vectors with zero mean and a scalar covariance
matrix, $\sigma^2 I$. In this case the maximum likelihood estimate of
$\Theta$ is given by the {\em orthogonal least squares fit} (OLSF),
which is based on the minimization of the function
\be
       {\cal F}(\Theta) =  \sum_{i=1}^n d_i^2
         \label{Fmain1}
\ee
where $d_i$ denotes the distance from the point $(x_i,y_i)$ to the
curve $P(x,y;\Theta)=0$.

Under these assumptions the OLSF is statistically optimal -- it
provides estimates of $\Theta$ whose covariance matrix attains its
Rao-Cramer lower bound \cite{CL03a,Ka96,Ka98}. The OLSF is widely used
in practice, especially when one fits simple curves such as lines or
circles. However, for more general curves the OLSF becomes intractable,
because the precise distance $d_i$ is hard to compute. In those cases
one resorts to various alternatives, and the most popular one is the
{\em algebraic fit} (AF) based on the minimization of
\be
       {\cal F}_{\rm a}(\Theta) =
       \sum_{i=1}^n w_i\, [P(x_i,y_i;\Theta)]^2
         \label{Fmain2}
\ee
where $w_i=w(x_i,y_i;\Theta)$ are suitably defined weights. The choice
of the weight function $w(x,y;\Theta)$ is important.
The AF is known \cite{CL03a} to provide a statistically optimal
estimate of $\Theta$ (in the sense that the covariance matrix will
attain its Rao-Cramer lower bound) if and only if the weight
function satisfies
\be
   w(x,y;\Theta) = a(\Theta) / \|\nabla P(x,y;\Theta)\|^2
     \label{wgrad}
\ee
for all points $x,y$ on the curve, i.e.\ such that
$P(x,y;\Theta)=0$. Here $\nabla P = (\partial P/\partial
x,\partial P/\partial y)$ is the gradient vector of the polynomial
$P$, and $a(\Theta)>0$ may be an arbitrary function of $\Theta$
(in practice, one simply sets $a(\Theta)=1$). Any other choice of
$w$ will result in the loss of accuracy, see \cite{CL03a}. We call
$w(x,y;\Theta)$ a {\em gradient weight function} if it satisfies
(\ref{wgrad}) for all $x,y$ on the curve $P(x,y;\Theta)=0$. The AF
(\ref{Fmain2}) with a gradient weight function $w(x,y;\Theta)$ is
commonly referred to as the {\em gradient weighted algebraic fit}
(GRAF). It was introduced in the mid-seventies \cite{Tu74} and
recently became standard for polynomial curve fitting, see, for
example, \cite{CBH01,LM00,Ta91}.

Even though the GRAF is much cheaper than the OLSF, it is still a
nonlinear problem requiring iterative methods. For example, in a
popular {\em reweight procedure} \cite{Sa82,Ta91} one uses the $k$-th
approximation $\Theta^{(k)}$ to compute the weights $w_i =
w(x_i,y_i;\Theta^{(k)})$ and then finds $\Theta^{(k+1)}$ by minimizing
(\ref{Fmain2}) regarding the just computed $w_i$'s as constants. Note
that if the parameters $\Theta$ are the coefficients of $P$, then
(\ref{Fmain2}), with fixed weights, becomes a quadratic function in
$\Theta$, and its minimum can be easily found. Another algorithm is
based on solving the equation $\nabla_{\Theta}{\cal F}_{\rm a}(\Theta)
= 0$, i.e.\
\be
   \sum P_i^2 \, \nabla_{\Theta} w_i +
   2 \sum w_i \, P_i \, \nabla_{\Theta} P_i = 0
      \label{weq}
\ee
for which various iterative schemes could be used. In the case of
fitting quadrics, for example, the most advanced algorithms are the
renormalization method \cite{Ka96}, the heteroscedastic
error-in-variables method \cite{LM00} and the fundamental numerical
scheme \cite{CBH01}. In all these algorithms, one needs to evaluate
${\cal O}(n)$ terms at each iteration. Therefore, the complexity of
those algorithms is ${\cal O}(kn)$, where $k$ is the number of
iterations. Moreover, each algorithm requires access to individual
coordinates $x_i,y_i$ of the data points at each iteration. These
difficulties can be sometimes avoided in a remarkable way, as we show
next.

Suppose we need to fit circles given by equation
$$
     P(x,y)=(x-a)^2+ (y-b)^2-R^2=0.
$$
Then we have
\be
   \|\nabla P(x,y;\Theta)\|^2 =
    4(x-a)^2 +4(y-b)^2\\
   = 4P(x,y) + 4R^2
     \label{4444}
\ee
hence $\|\nabla P(x,y;\Theta)\|^2 = 4R^2$ for all the points
$(x,y)$ lying on the circle $P(x,y)=0$, and we can set
$w(x,y;\Theta) = 1/R^2$. Therefore
\begin{eqnarray}
    {\cal F}_{\rm a}(a,b,R) &=&
       \sum_{i=1}^n R^{-2} \left[x_i^2+y_i^2-2ax_i-2by_i+
       a^2+b^2-R^2\right]^2\nonumber\\
       &=& R^{-2}[z_1+az_2+bz_3+a^2z_4+b^2z_5+abz_6
       +cz_7+ac z_8+bc z_9+ c^2n]
         \label{FmainC}
\end{eqnarray}
where we denoted $c=a^2+b^2-R^2$ for brevity, and
$$
   z_1=\sum (x_i^2+y_i^2)^2,\ z_2=-4\sum x_i(x_i^2+y_i^2),\ldots
$$
are some expressions involving $x_i$ and $y_i$ only.

The minimization of (\ref{FmainC}) is still a nonlinear problem
requiring iterative methods \cite{CO84,CL02,Pr87}, but it has
obvious advantages over the reweight procedure described above and
other generic methods for solving the equation (\ref{weq}). First
of all, the values of $z_1,\ldots,z_9$ only need to be computed
once, and then the cost of minimization of (\ref{FmainC}) will not
depend on $n$ anymore. Thus, the complexity of this algorithm is
${\cal O}(n) + {\cal O}(k)$, where ${\cal O}(n)$ is the cost of
evaluation of $z_1,\ldots,z_9$ and ${\cal O}(k)$ is the cost of
some $k$ iterations spent on the subsequent minimization of ${\cal
F}_{\rm a}(a,b,R)$. Moreover, once the values of $z_1,\ldots,z_9$
are computed and stored, the coordinates $x_i,y_i$ can be
destroyed. Practically, $z_1,\ldots,z_9$ can be computed
``on-line'', when the data are collected. The minimization
procedure per se can be implemented ``off-line'', without storage
of (or access to) the data points. The quantities $z_1,\ldots,z_9$
here play the role of sufficient statistics.

Inspired by the above example, we might say that the problem of fitting
a polynomial curve $P(x,y;\Theta)=0$ {\em admits a reduction of
complexity} if there are $\ell$ functions
$z_j(x_1,y_1,\ldots,x_n,y_n)$, $1\leq j\leq\ell$, with $\ell $ being
independent of $n$ and $\Theta$, and a gradient weight function
$w(x,y;\Theta)$ such that
\be
       {\cal F}_{\rm a} = f(z_1,\ldots,z_{\ell};\Theta)
         \label{Fzz}
\ee
i.e.\ ${\cal F}_{\rm a}$ is a function of $z_1,\ldots,z_{\ell}$ and
$\Theta$ only.

This definition does not suggest how to find the functions
$z_1,\ldots,z_{\ell}$ in practical terms, though. Since ${\cal F}_{\rm
a}$ is given by (\ref{Fmain2}) with $P(x_i,y_i;\Theta)$ being a
polynomial in $x_i,y_i$, then the most natural (if not the only) way to
construct the functions $z_1,\ldots,z_{\ell}$ is to express the
gradient weight function (\ref{wgrad}) in the form
\be
       w(x,y;\Theta) = \sum_{k=1}^K C_k(\Theta)\, D_k(x,y)
         \label{wCD}
\ee
where $C_k$ are functions of the parameter vector $\Theta$ alone, and
$D_k$ are functions of $x$ and $y$ only (here the number of terms, $K$,
must be independent of $\Theta$). Indeed, suppose that the
representation (\ref{wCD}) is found. Since $P^2$ is a polynomial in
$x,y$, we can expand it as
$$
    P^2(x,y) = \sum_{p,q} c_{p,q}x^py^q
$$
where $c_{p,q} = c_{p,q} (\Theta)$ denote its coefficients. Now the
function ${\cal F}_{\rm a}$ can be evaluated as
\begin{eqnarray*}
   {\cal F}_{\rm a} &=& \sum_{k=1}^K\sum_{p,q}
   C_k(\Theta)c_{p,q}(\Theta)
   \sum_{i=1}^n x_i^py_i^qD_k(x_i,y_i) \\
   &=& \sum_{k=1}^K\sum_{p,q}
   C_k(\Theta)c_{p,q}(\Theta)
   z_{k,p,q}
\end{eqnarray*}
where
$$
   z_{k,p,q} = \sum_{i=1}^n x_i^py_i^qD_k(x_i,y_i)
$$
The values of $z_{k,p,q}$ depend on the data $x_i,y_i$ only, hence we
obtain the desired representation (\ref{Fzz}). Therefore, (\ref{wCD})
implies (\ref{Fzz}). We believe that the converse is also true, i.e.\
the conditions (\ref{Fzz}) and (\ref{wCD}) are actually equivalent, but
we do not attempt to prove that.

Motivated by the above considerations, we adopt the following
definition: the problem of fitting a polynomial curve $P(x,y;\Theta)=0$
{\em admits a reduction of complexity} if the gradient weight function
(\ref{wgrad}) can be expressed in the form (\ref{wCD}).

As we have seen, the problem of fitting circles admits a reduction
of complexity (and so does the simpler problem of fitting lines).
Now if the problem of fitting ellipses and/or hyperbolas admitted
a reduction of complexity as defined above, we would be able to
dramatically improve the known GRAF algorithms
\cite{CBH01,Ka96,LM00}. Unfortunately, this is impossible -- there
are deep mathematical reasons which prevent a reduction of
complexity in the case of ellipses, hyperbolas, and parabolas.

In this paper we find general conditions on the polynomial
$P(x,y;\Theta)$ under which the problem of fitting the curve
$P(x,y;\Theta)=0$ allows a reduction of complexity. It turns out
that lines and circles satisfy these conditions, but ellipses,
hyperbolas, and parabolas do not. Our results thus demonstrate (in
a rigorous mathematical way) that fitting noncircular conics is an
intrinsically more complicated problem than fitting circles or
lines.

For convenience, let us denote
$$
   Q(x,y;\Theta) :=
   \|\nabla P(x,y;\Theta)\|^2 =
   (\partial P/\partial x)^2+(\partial P/\partial y)^2
$$
Clearly, $Q(x,y;\Theta)$ is itself a polynomial in $x$ and $y$. Our
subsequent arguments will involve some facts from complex analysis. We
will treat $x$ and $y$ as {\em complex}, rather than {\em real},
variables.

\medskip\noindent{\bf Theorem}. {\em The problem of fitting curves
$P(x,y;\Theta)=0$ admits a reduction of complexity (as defined
above) under the condition that the system of polynomial
equations}
\be
  \left \{
  \begin{array}{c}
  P(x,y) = 0\\
  Q(x,y) = 0
  \end{array} \right .
  \label{PQ0}
\ee
{\em has no solutions, real or complex.}
\medskip

Before we prove our theorem, we shall show how to use it. For the
problem of fitting circles, we have already computed $Q=4P+4R^2$, see
(\ref{4444}), hence the system (\ref{PQ0}) has indeed no solutions for
nondegenerate circles (for which $R\neq 0$).

When using the theorem, the following {\em invariance} property
will be helpful. Let $(x,y)\mapsto (\tilde{x},\tilde{y})$ be a
transformation of the $xy$ plane that is a composition of
translations, rotations, mirror reflections and similarities (the
latter are defined by $(x,y)\mapsto (cx,cy)$ for some $c\neq 0$).
Denote by $\tilde{P}(\tilde{x},\tilde{y})$ the polynomial $P$ in
the new coordinates $\tilde{x},\tilde{y}$. Then the system
(\ref{PQ0}) has a solution (real or complex) if and only if the
corresponding system
$$
  \left \{
  \begin{array}{c}
  \tilde{P}(\tilde{x},\tilde{y}) = 0\\
  \tilde{Q}(\tilde{x},\tilde{y}) = 0
  \end{array} \right .
$$
has a solution, real or complex. Here $\tilde{Q} = \|\nabla
\tilde{P}\|^2$. This simple fact, which can be verified directly
by the reader, allows us to simplify the polynomial $P(x,y)$
before applying the theorem.

Consider the problem of fitting ellipses and hyperbolas. By using
a translation and rotation of the $xy$ plane we can always reduce
the polynomial $P$ to a canonical form $ax^2+by^2+c=0$ (with
$a\neq b$ and $abc\neq 0$). Then $Q=4a^2x^2+4b^2y^2$ and we arrive
at a system of equations
$$
  \left \{
  \begin{array}{c}
   ax^2+by^2+c = 0\\
   a^2x^2+b^2y^2 = 0
  \end{array} \right .
$$
It is easy to see that it always has a solution
$$
   x=\pm\sqrt{\frac{bc}{a(a-b)}},\quad
   y=\pm\sqrt{-\frac{ac}{b(a-b)}}
$$
(note that $x$ or $y$ may be an imaginary number, which is allowed
by our theorem). Therefore, the problem does not admit a reduction
of complexity.

If our curve is a parabola, then we can use its canonical equation
$y=cx^2$ for $c>0$, hence $P=y-cx^2$ and $Q=4c^2x^2+1$. Here again we
have a common zero of $P$ and $Q$ at the point $x={\bf i}/2c$ and
$y=-1/4c$. Thus, no conic sections (except circles) satisfy the
conditions of our theorem.


We now prove our theorem. Since $w(x,y;\Theta)$ must be a gradient
weight function, the requirement (\ref{wCD}) is equivalent to
\be
     \frac{1}{Q(x,y)} = \sum_{k=1}^K C_k(\Theta)\, D_k(x,y)
     \ \ \ \ \ \ {\rm whenever}\ \ \ \
     P(x,y)=0
       \label{QU}
\ee
(here we incorporated the factor $a(\Theta)$ into the coefficients
$C_k(\Theta)$, for convenience). We emphasize that the left
identity in (\ref{QU}) does not have to hold on the entire $xy$
plane, it only has to hold {\em on the curve} $P(x,y)=0$. If we
denote that curve by $\cal L$, then (\ref{QU}) can be restated as
\be
     \frac{1}{Q(x,y)} = \sum_{k=1}^K C_k(\Theta)\, D_k(x,y)
     \ \ \ \ \ \ {\rm whenever}\ \ \ \
     (x,y)\in{\cal L}
       \label{QU1}
\ee

The functions $D_k(x,y)$ in (\ref{QU}) cannot be arbitrary, they
must be easily computable, i.e.\ available in the machine
arithmetics. That is, they must be combinations of elementary
functions -- polynomials, exponentials, logarithms, trigonometric
functions, etc. In that case $D_k(x,y)$ are analytic functions of
$x$ and $y$. Therefore, they have analytic extensions to the
two-dimensional complex plane $\IC^2$. We note that they do not
need be {\em entire functions}, i.e.\ analytic everywhere in
$\IC^2$, they may have some singularities. For example, the
function $(1+x^2+y^2)^{-1}$ is analytic in $\IR^2$ but has
singularities in $\IC^2$, e.g.\ the point $x={\bf i}$ and $y=0$ is
its singularity. Also, those extensions maybe multivalued
functions (examples are $\ln x$ or $\sqrt{x}$).

Now, the following function will also be analytic in $\IC^2$:
$$
  G(x,y) = 1 - Q(x,y)\sum_{k=1}^K C_k(\Theta)\, D_k(x,y)
$$
since it is a combination of analytic functions. By (\ref{QU1}),
it vanishes on the curve $\cal L$ in the real $xy$ plane. Consider
the subset ${\cal Z}\subset\IC^2$ defined by the equation
$P(x,y)=0$, where $x$ and $y$ are treated as complex variables.
Note that $\cal L$ is a curve on the two-dimensional manifold
$\cal Z$. We will prove that the function $G(x,y)$ vanishes on the
entire $\cal Z$.

We can assume that $P(x,y)$ is an irreducible polynomial
(otherwise we can apply our argument to each irreducible factor of
$P$). Then $\cal Z$ is an algebraic variety, hence it admits a
complex parametrization (a complex coordinate, $z$), and the
restriction of the function $G$ onto $\cal Z$ will be an analytic
function of $z$. It is known in complex analysis that if an
analytic function $G(z)$, $z\in\IC$, vanishes on a one-dimensional
curve in $\IC$, then it is identically zero on $\IC$, hence
$G(z)\equiv 0$ for all $z\in\IC$. In our case the curve on which
$G$ vanishes is $\cal L$ (and we assume, of course, that it is a
nondegenerate curve for all the relevant values of the parameter
$\Theta$). Hence, $G$ vanishes on the entire $\cal Z$, and
therefore
\be
     G(x,y)=0
     \ \ \ \ \ \ {\rm whenever}\ \ \ \
     (x,y)\in{\cal Z}
       \label{GP}
\ee
On the other hand, if the system of equations (\ref{PQ0}) has a
complex solution $(x,y)$, then (\ref{GP}) would be impossible,
since any solution of (\ref{PQ0}) lies on the manifold $\cal Z$
(because $P(x,y) = 0$), and at the same time $Q(x,y)=0$ implies
$G(x,y) = 1$. Therefore, if the system (\ref{PQ0}) has a solution
(real or complex), then the representation (\ref{wCD}) cannot
possibly exist.


It remains to show that if the system (\ref{PQ0}) has no solutions,
then the representation (\ref{wCD}) is possible, and hence our problem
indeed admits a reduction of complexity. Assuming that (\ref{PQ0}) has
no solutions, we will construct the representation (\ref{wCD}) in the
simplest, polynomial form:
\be
       w(x,y;\Theta) = \sum_{p,q} w_{p,q}(\Theta)\, x^py^q
         \label{wmn}
\ee
the degree of this polynomial being independent of the parameter
$\Theta$. Consider a polynomial equation
\be
    P(x,y)\, U(x,y) + Q(x,y)\, W(x,y) = 1
      \label{UW}
\ee
where $U(x,y)$ and $W(x,y)$ are unknown polynomials. A classical
mathematical theorem, Hilbert's Nullstellensatz \cite{ZS}, says
that the equation (\ref{UW}) has polynomial solutions $U(x,y)$ and
$W(x,y)$ if and only if $P(x,y)$ and $Q(x,y)$ have no common
zeroes in $\IC^2$, i.e.\ whenever the system (\ref{PQ0}) has no
complex solutions, which is exactly what we have assumed. Note
that since $P$ and $Q$ depend on $\Theta$, then so do $U$ and $W$,
but we suppress this dependence in the equation (\ref{UW}).

Now the polynomial $W(x,y)$ solving (\ref{UW}) gives us the weight
function $w(x,y;\Theta)=W(x,y)$, and it is easy to see that
$$
     W(x,y) = 1/Q(x,y)
     \ \ \ \ \ \ {\rm whenever}\ \ \ \
     P(x,y) = 0
$$
Technically, the theorem is proved, but we make a further
practical remark. Suppose we know that the system (\ref{PQ0}) has
no solutions, so that the problem admits a reduction of
complexity. In this case we need to find the polynomial $W(x,y)$
solving (\ref{UW}) in an explicit form, in order to determine the
weight function $w(x,y;\Theta)$. To this end we describe a finite
and relatively simple algorithm for computing the coefficients
$w_{pq}$ of the polynomial $W$. We substitute the expansions
$$
       W(x,y) = \sum_{p,q} w_{p,q}\, x_i^py_i^q
       \ \ \ \ \ {\rm and}\ \ \ \ \
       U(x,y) = \sum_{p,q} u_{p,q}\, x_i^py_i^q
$$
into the identity (\ref{UW}) and then equate the terms on the left hand
side and those on the right hand side with the same degrees of the
variables $x,y$. This gives a linear system of equations for the
unknown coefficients $w_{pq}$ and $u_{pq}$. This might be a large
system (its size depends on the degrees of $U$ and $W$), but it is a
linear system whose solution can be found by routine matrix methods. If
the assumed degrees of $U$ and $V$ are high enough, then the above
system is always solvable by the so called {\em effective
Nullstellensatz}, see \cite{S67}. By solving that system we can obtain
explicit formulas for the coefficients $w_{pq}$ and $u_{pq}$. In fact,
we only need the coefficients of $W$, not $U$. Lastly, we remark that
those coefficients will be rational functions of the coefficients of
the polynomial $P(x,y)$, hence they will be easily computable.

\noindent{\bf Acknowledgement}. N. Chernov is partially supported by
NSF grant DMS-0098788 and N.~Sim\'{a}nyi is partially supported by NSF
grant DMS-0098773.


\begin{thebibliography}{99}

\bibitem{BC86} M. Berman and D. Culpin,
    The statistical behaviour of some least squares estimators of the centre and radius of a
    circle,  {\em J. R. Statist. Soc. B}, {\bf 48}, 1986, 183--196.

\bibitem{CO84} N. I. Chernov and G. A. Ososkov,
    Effective algorithms for circle fitting,
    {\em Comp. Phys. Comm.} {\bf 33}, 1984, 329--333.

\bibitem{CL02} N. Chernov and C. Lesort,
    {\rm Fitting circles and lines by least squares: theory and experiment},
    preprint, available at http://www.math.uab.edu/cl/cl1

\bibitem{CL03a} N. Chernov and C. Lesort,
    {\rm Statistical efficiency of curve fitting algorithms},
    preprint, available at http://www.math.uab.edu/cl/cl2

\bibitem{CBH01} W. Chojnacki, M. J. Brooks, and A. van den Hengel,
    {\rm Rationalising the renormalisation method of Kanatani},
    {\em J. Math. Imaging \& Vision}, {\bf 14}, 2001, 21--38.

\bibitem{GGS94} W. Gander, G. H. Golub, and R. Strebel,
    {\rm Least squares fitting of circles and ellipses},
    {\em BIT} {\bf 34}, 1994, 558--578.

\bibitem{Ka96} K. Kanatani,
    {\em Statistical Optimization for Geometric Computation: Theory and Practice},
    Elsevier Science, Amsterdam, 1996.

\bibitem{Ka98} K. Kanatani,
    {\rm Cramer-Rao lower bounds for curve fitting},
    {\em Graph. Models Image Proc.} {\bf 60}, 1998, 93--99.

\bibitem{LM00} Y. Leedan and P. Meer,
    {\rm Heteroscedastic regression in computer vision: Problems with bilinear
    constraint},
    {\em Intern. J. Comp. Vision}, {\bf 37}, 2000, 127--150.

\bibitem{Pr87} V. Pratt,
    {\rm Direct least-squares fitting of algebraic surfaces},
    {\em Computer Graphics} {\bf 21}, 1987, 145--152.

\bibitem{Sa82} P. D. Sampson,
    {\rm Fitting conic sections to very scattered data:
    an iterative refinement of the Bookstein algorithm},
    {\em Comp. Graphics Image Proc.} {\bf 18}, 1982, 97--108.

\bibitem{S67} J. R. Shoenfield, {\em Mathematical logic}, Reading, Mass.,
Addison-Wesley, 1967, p. 100, Ex. 18 (e).

\bibitem{Ta91} G. Taubin,
    {\rm Estimation Of Planar Curves, Surfaces And Nonplanar
    Space Curves Defined By Implicit Equations,
    With Applications To Edge And Range Image Segmentation},
    {\em IEEE Transactions on Pattern Analysis and Machine
    Intelligence},  {\bf 13}, 1991, 1115--1138.

\bibitem{Tu74} K. Turner, {\em Computer perception of curved
objects using a television camera}, Ph.D.\ Thesis, Dept.\ of Machine
Intelligence, University of Edinburgh, 1974.

\bibitem{ZS} O. Zariski and P. Samuel, {\em Commutative algebra}, Vol. 2.
Princeton, N.J., Van Nostrand [1958-60], p. 164.

\end{thebibliography}
\end{document}